\documentclass[aps,prc,reprint,groupedaddress,twocolumn,longbibliography]{revtex4-2}
\usepackage{graphicx} 
\usepackage{subfigure}
\usepackage{amsmath}
\usepackage{physics}
\usepackage{hyperref}
\usepackage{indentfirst}
\usepackage{geometry}
\usepackage{dcolumn}
\usepackage{bm}
\usepackage{etoolbox}

\newcommand{\fig}[1]{Fig.~\ref{#1}}
\hypersetup{colorlinks=true, citecolor=blue, urlcolor=blue, linkcolor=blue}

\begin{document}

    \title{\textbf{Energy dependence of rescattering effect on vector mesons spin alignment at RHIC}}
    \author{Zihan Liu}
    \author{Ziyang Li}
    \author{Wangmei Zha}
    \author{Zebo Tang}
    \email{zbtang@ustc.edu.cn (Corresponding author)}
    \affiliation{State Key Laboratory of Particle Detection and Electronics, University of Science and Technology of China, Hefei 230026, China}
    \date{\today}
    
    \begin{abstract}
    Spin alignment of vector mesons in heavy-ion collisions provides a novel probe of quark polarization and hadronization mechanism in quark-gluon plasma. Hadronic rescattering may affect the measured spin alignment of vector mesons due to non-uniform rescattering probability in non-central heavy-ion collisions. Using the UrQMD model, we systematically investigated the hadronic rescattering effect on the measurement of $\rho_{00}$, the spin alignment parameter, for $K^{*0}$, $\phi$, and $\rho^{0}$ mesons in Au+Au collisions at $\sqrt{s_\mathrm{NN}}$ = 7.7 - 200 GeV. Our results reveal that the measurable $\rho_{00} - 1/3$ remains unaffected for $\phi$, while shows significantly negative (positive) deviations for $K^{*0}$ and $\rho^0$ with respect to the reaction (production) plane. Quantitatively, the maximum deviation reaches $-0.0056$ ($0.0268$) for $K^{*0}$ and $-0.0122$ ($0.0414$) for $\rho^{0}$ with respect to the reaction (production) plane. Notably, the deviations in $\rho_{00}$ for both $K^{*0}$ and $\rho^{0}$ increase monotonically with increasing collision energy. These findings underscore the critical necessity of accounting for rescattering effects when interpreting spin alignment measurements of short-lived vector mesons in heavy-ion collisions.
    
    \end{abstract}

    \maketitle
	
	\section{\label{section:1}Introduction}	
	
	Heavy ions can be accelerated to velocities approaching the speed of light and collide from opposite directions, creating a hot and dense medium of deconfined quarks and gluons, known as the quark-gluon plasma (QGP)~\cite{QGP1,QGP2}. It provides an ideal opportunity to investigate phenomena in non-perturbative quantum chromodynamics (QCD). In non-central heavy-ion collisions, a large orbital angular momentum $(\approx10^{4}\,\hbar)$~\cite{LargeL} and magnetic field $(\approx10^{14}\,\mathrm{T})$~\cite{LargeB} are also expected. While the magnetic field is short-lived, such a large angular momentum is conserved and could affect the system throughout its evolution. Due to spin-orbital coupling, this angular momentum can polarize quark spins along its direction—a phenomenon known as global polarization~\cite{polarization1,polarization1_erratum,polarization2,polarization3}. Consequently, the spin of hadrons composed of these polarized quarks may also be polarized. In 2017, the STAR Collaboration first reported the discovery of a significant global polarization of the $\Lambda(\bar{\Lambda})$ hyperons in Au+Au collisions~\cite{lambda_polarization}. According to the flavor-spin wave function, the polarization of $\Lambda(\bar{\Lambda})$ hyperon is carried solely by the strange quark $s(\bar{s})$, indicating global polarization of $s(\bar{s})$ quark. Since strange quarks can also form vector mesons such as $K^{*0}(892)$ and $\phi(1020)$, their spin could also be influenced. Therefore, this quark-level polarization along a global direction could be reflected in the spin alignment of vector mesons and measured in experiments.
    
	The spin state of a vector meson is characterized by a $3\times3$ Hermitian spin-density matrix $\rho$ with unit trace. The diagonal elements $\rho_{-1-1}, \rho_{00}$, and $\rho_{11}$ represent the probabilities of vector meson's spin being $-1$, $0$, and $1$, respectively. However, $\rho_{-1-1}$ and $\rho_{11}$ cannot be measured separately in two-body decays, leaving $\rho_{00}$ as the only accessible independent element. The value of $\rho_{00}$ can be determined from the angular distribution of decay daughters as follows~\cite{rho00_derivation},
	\begin{equation}
		\centering
        \label{eq1}
		\dv{N}{\cos \theta^*}\propto\qty[(1-\rho_{00})+(3\rho_{00}-1)\cos^2\theta^*]
	\end{equation}
	where $\theta^*$ denotes the angle between the quantization axis and the momentum direction of a decay daughter in the rest frame of its parent vector meson. A value of $\rho_{00}=1/3$ corresponds to no spin alignment, while any deviation from 1/3 indicates the presence of spin alignment. 
	
	The ALICE Collaboration first reported evidence of spin alignment for $K^{*0}$ and $\phi$ mesons with respect to both the event plane and the production plane in semi-central Pb + Pb collisions at $\sqrt{s_\mathrm{NN}} = 2.76$ TeV~\cite{kstarphi_sa_alice}. The measured $\rho_{00}$ values for $K^{*0}$ and $\phi$ are found to be less than 1/3 at low transverse momentum ($p_\mathrm{T}<2$ GeV/$c$) with statistical significance of $3\sigma$ and $2\sigma$, respectively. These findings are consistent with theoretical expectations and suggest that vector mesons composed of polarized quarks can exhibit patterns of spin alignment. Subsequently, the STAR Collaboration measured $K^{*0}$ and $\phi$ spin alignment in semi-central Au+Au collisions in beam energy scan program phase I (BES-I) energies~\cite{kstarphi_sa_star}. Their results indicate that $\rho_{00}$ for $K^{*0}$ is consistent with 1/3 at $\sqrt{s_\mathrm{NN}} \leq 54.4$ GeV, while $\rho_{00}$ for $\phi$ exceeds 1/3 with a significance of $7.4\sigma$ at $\sqrt{s_\mathrm{NN}} \leq 62$ GeV, which contradicts the findings in 2.76 TeV Pb + Pb collisions. Moreover, $\rho_{00}$ for $\phi$ exhibits a decreasing trend with increasing collision energy. The analyses of $\rho_{00}$ for $K^{*0}$ and $\phi$ in beam energy scan program phase II (BES-II) are ongoing with the aim of more precise measurements~\cite{sqm2024}. These observations cannot be fully explained by conventional polarization mechanisms; however, they may be qualitatively understood within a $\phi$ meson field framework, in which the strong electric component of the $\phi$ meson field polarizes the $s(\bar{s})$ quarks~\cite{phi_meson_field1, phi_meson_field1_err, phi_meson_field2}. This theory is analogous to how an electric field polarizes a quark through spin-orbit coupling, but with much stronger magnitude due to its strong interaction. However, the discrepancy in $\rho_{00}$ for vector mesons between ALICE and STAR measurements remains not well understood. Further quantitative explanations are still required, and it is essential to explore additional possible contributions to the spin alignment measurements in order to understand the polarization mechanism.

    One major difference between the $K^{*0}$ and $\phi$ mesons lies in their lifetime. The $K^{*0}$ is a short-lived resonance with a lifetime of approximately 4.2 fm/$c$, which is comparable to the time span between chemical freeze-out (CFO) and kinetic freeze-out (KFO)~\cite{pdg, time_cfokfo}. In contrast, the lifetime of $\phi$ is almost ten times longer than $K^{*0}$. As a result, $K^{*0}$ tends to decay during the hadronic phase and is thus more sensitive to hadronic interactions, while $\phi$ typically survives beyond this phase. In experiments, these vector mesons are commonly reconstructed via their strong decay channels. Short-lived resonances may decay within the dense hadronic medium created in heavy-ion collisions, where their decay daughters are likely to experience hadronic rescattering. Such rescattering can either alter the momenta of decay daughters or lead to their absorption, thereby hindering the reconstruction of the parent resonances. The rescattering effect on the measurement of $K^{*0}$ production has been observed at various energies and collision systems~\cite{kstarbes1, kstar62.4_200gev, kstar_phi2.76tev,kstar_phi2.76tev_rescattering}. In contrast, measurements of $\phi$ production suggest that $\phi$ is barely affected by rescattering due to its longer lifetime~\cite{phi200gev, phi62.4_200gev, kstar_phi2.76tev, kstar_phi2.76tev_rescattering}. Moreover, in non-central heavy-ion collisions, the medium exhibits spatial anisotropy, leading to a non-uniform density profile. This would result in a non-uniform angular distribution of the rescattering probability and an angular dependence of the reconstructable parents. Since the spin alignment of vector mesons is extracted from the angular distribution of decay daughters, rescattering can produce artificial spin alignment for short-lived resonances and thereby affect the measured value of their $\rho_{00}$.
    
    Previous studies have investigated the rescattering effect on $K^{*0}$ spin alignment at $\sqrt{s_\mathrm{NN}} = 200$ GeV, finding it to be significant compared to various theoretical sources~\cite{kstar_spin_rescatterin,AMPT_result}. However, the rescattering effect on $K^{*0}$ spin alignment in BES energies has yet to be studied, and similar studies are also needed for other vector mesons, such as $\phi$ and $\rho^{0}$. Investigating the rescattering effect on the measurement of spin alignment for $K^{*0}$ and $\phi$ at BES energies is crucial to accurately interpret the measured $\rho_{00}$ values. In addition, compared to $K^{*0}$ and $\phi$, the spin alignment of $\rho^{0}$ would provide a crucial component in the background estimation for chiral magnetic effect (CME) measurements involving pions~\cite{rho_CME}. The $\rho^0$ meson, with an even shorter lifetime, is more susceptible to rescattering in the hadronic phase. The suppression of $\rho^0$ in central Pb + Pb collisions with respect to $pp$ collisions has been observed and attributed to this effect on the measurement of its production~\cite{rho_2.76tev_PbPb}. Thus, it is essential to investigate the rescattering effect on the measurement of spin alignment of $\rho^0$ mesons as well.
    
    
    
	
	In this paper, the rescattering effect on the spin alignment of $K^{*0}$, $\phi$ and $\rho^{0}$ is studied with the ultrarelativistic quantum molecular dynamics (UrQMD) model at mid-rapidity ($\abs{y} < 1.0$) in Au+Au collisions at $\sqrt{s_\mathrm{NN}} = 7.7$, $9.2$, $11.5$, $14.5$, $17.3$, $19.6$, $27$, $39$, $62.4$ and $200$ GeV, ranging from BES energies to the RHIC top energy. The paper is organized as follows. Section \ref{section:2} introduces the methodology, Section \ref{section:3} presents the results and discussions, and Section \ref{section:4} summarizes the results.
	
	\section{\label{section:2}Methodology}	
	
	\subsection{UrQMD model}
	
	This study employs the UrQMD model to investigate the rescattering effect on the measurement of spin alignment for vector mesons. UrQMD is a microscopic transport model for modeling heavy-ion collisions over a broad energy range, from the SIS to the LHC. It is based on the covariant propagation of strings and hadrons to simulate binary scatterings, color string formation, and resonance decay~\cite{urqmd1,urqmd2}. The model includes the full set of established hadrons with more than 50 baryon species and 40 meson species. The interaction cross sections are taken either from experimental data or calculated by using detailed balance principle. In the UrQMD model, the initial orbital angular momentum is not explicitly propagated. Hence, the spin alignment for the initially generated vector mesons is not expected, and any deviation of $\rho_{00}$ from 1/3 is attributed to the hadronic interactions, i.e. the rescattering effect.
    
    To reconstruct a resonance through its two-body decay channel in experiment, one needs to pair all unlike-sign candidates in the same event first and calculate their invariant masses with a substantial combinatorial background. The resonance signal, with a typical signal-to-background ratio on the order of $1/100$, could be extracted after subtracting the combinatorial background~\cite{kstarbes1,kstar62.4_200gev}. In UrQMD, there is no need to pair all unlike-sign candidates, one can track the resonances and their decay daughters directly due to one fantastic feature of this model, its history file. The history file records both the phase-space information of all output particles and every interaction, including hadron scattering and resonance decay.
	
	\subsection{Resonances reconstruction}
	
	In this study, three types of vector mesons, $\rho^0, K^{*0}$ (as well as its antiparticle $\overline{K^{*0}}$) and $\phi$, are reconstructed via their strong decay channels. Detailed information is shown in Table \ref{tab:1}. Both $K^{*0}$ and $\overline{K^{*0}}$ are used in this study; henceforth, we denote them by $K^{*0}$ unless specified.
	
	\begin{table}[htbp]
		\centering
		\caption{\label{tab:1}Lifetime, decay channels and branching ratios used in the reconstruction of vector mesons~\cite{pdg}}		  
		\begin{tabular}{cccc}
			\hline\hline\noalign{\smallskip}	
			Meson & Lifetime (fm/$c$) & Decay channel & Branching Ratio\\
			\noalign{\smallskip}\hline\noalign{\smallskip}
			$\rho^0$ & 1.3  & $\rho^0\rightarrow\pi^+\pi^-$ & $100\%$\\
			$K^{*0}$ & 4.2 & $K^{*0}\rightarrow K^\pm\pi^\mp$ & $66.6\%$\\
			$\phi$ & 46.3 & $\phi\rightarrow K^+K^-$ & $49.1\%$ \\
			\noalign{\smallskip}\hline\hline
		\end{tabular}
	\end{table}
	
	We illustrate the reconstruction method using $K^{*0}$ as an example. First, all $K^{*0}\rightarrow K\pi$ decays with $K^{*0}$'s rapidity lays in $\abs{y}<1.0$ are identified from the collision history file. The decay daughters are then tracked throughout the whole evolution history to determine whether they undergo elastic or inelastic scattering. A $K^{*0}$ is considered reconstructable if neither of its decay daughters participates in any subsequent collisions. In contrast, a $K^{*0}$ is labeled unreconstructable if at least one of its decay daughters undergoes elastic or inelastic scattering with other particles during the hadronic evolution. Such scattering can alter the momenta of the daughters or even completely absorb them, thereby preventing the reconstruction of the parent meson. For all $K^{*0}$, they include both reconstructable and unreconstructable $K^{*0}$, without considering the rescattering of their decay products. The rescattering effect on the vector mesons is therefore estimated by comparing reconstructable $K^{*0}$ and all $K^{*0}$. Previous studies have shown that the reconstruction method of $K^{*0}$ generated in UrQMD works well~\cite{kstar_spin_rescatterin,kstarbes1_urqmd}. A similar approach is adopted to reconstruct $\phi$ and $\rho^0$ to estimate the rescattering effect on their spin alignment as well.
	
	\subsection{\texorpdfstring{$\rho_{00}$}{} measurement}

    \begin{figure}[htbp]
        \includegraphics[scale=0.4]{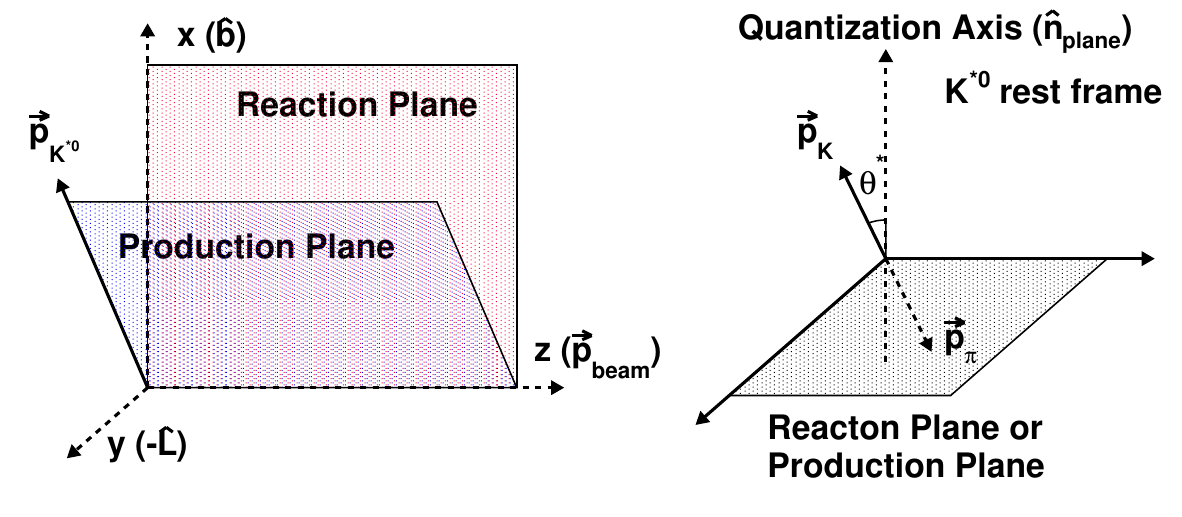}
        \caption{\label{fig:RP_PP}A schematic illustration of the reaction plane and the production plane, using $K^{*0}$ as an example. The vectors for $K^{*0}$ and $K, \pi$ represent their momenta, $\theta^*$ is the polar angle between the quantization axis and the momentum direction of a daughter in the rest frame of the decay.}        
    \end{figure}
	
	Two distinct reference planes are typically used in the analysis of the spin alignment of vector mesons: the reaction plane and the production plane~\cite{kstarphi_sa_star,kstarphi_sa_alice,kstar_spin_rescatterin}. A schematic illustration of these two planes is shown in Fig.~\ref{fig:RP_PP}. The reaction plane is defined by the cross product of the beam direction and the impact parameter. The value of $\rho_{00}$ measured with respect to reaction plane is referred to as the global spin alignment. The production plane, which is another popular choice of reference frames, is defined by the cross product of the beam direction and momentum of the vector meson. In this study, the spin alignment of vector mesons is investigated with respect to both the reaction and production plane, which can be readily determined in UrQMD. The angular distribution of decay daughters is fitted using Eq. \ref{eq1} for all and reconstructable vector mesons, respectively. The fitting procedure follows that described in Ref.~\cite{kstar_spin_rescatterin}.
	
	\section{\label{section:3}Results and Discussions}	
	
	\subsection{Particle yield ratios}	

    \begin{figure}[htbp]
		\centering
		\includegraphics[scale=0.4]{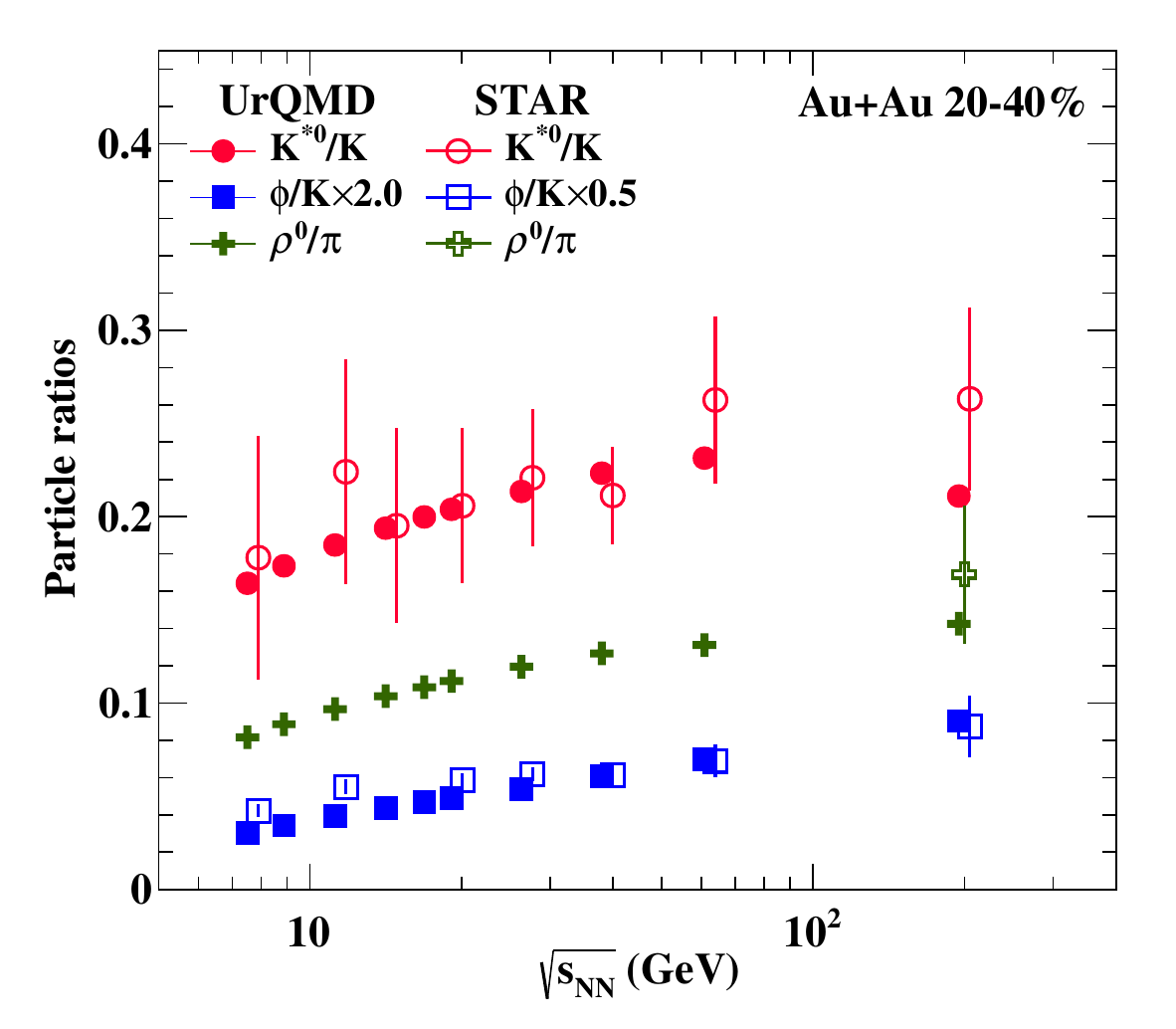}
		\caption{\label{fig:ratio_sNN}The yield ratios of $K^{*0}/K$, $\phi/K$ and $\rho^{0}/\pi$ at mid-rapidity as a function of $\sqrt{s_\mathrm{NN}}$ in $20-40\%$ Au+Au collisions from UrQMD at $\sqrt{s_\mathrm{NN}} = 7.7$, $9.1$, $11.5$, $14.5$, $17.3$, $19.6$, $27$, $39$, $62.4$ and $200$ GeV. The results are compared with previous STAR measurements~\cite{kstarbes1, kstar62.4_200gev, phi62.4_200gev, phi200gev, rho_ppAuAu_star}, whose vertical bars represent summed errors. At $\sqrt{s_\mathrm{NN}}=$ 200 GeV, $K^{*0}/K$ ratio is measured in $10-40\%$ Au+Au collisions and $\rho^{0}/\pi$ ratio is measured in $40-80\%$ Au+Au collisions both by UrQMD model and by STAR.}		
	\end{figure}

\begin{figure}[htbp]
		\centering
        \begin{minipage}{0.9\linewidth}
            \centering
            \includegraphics[scale=0.4]{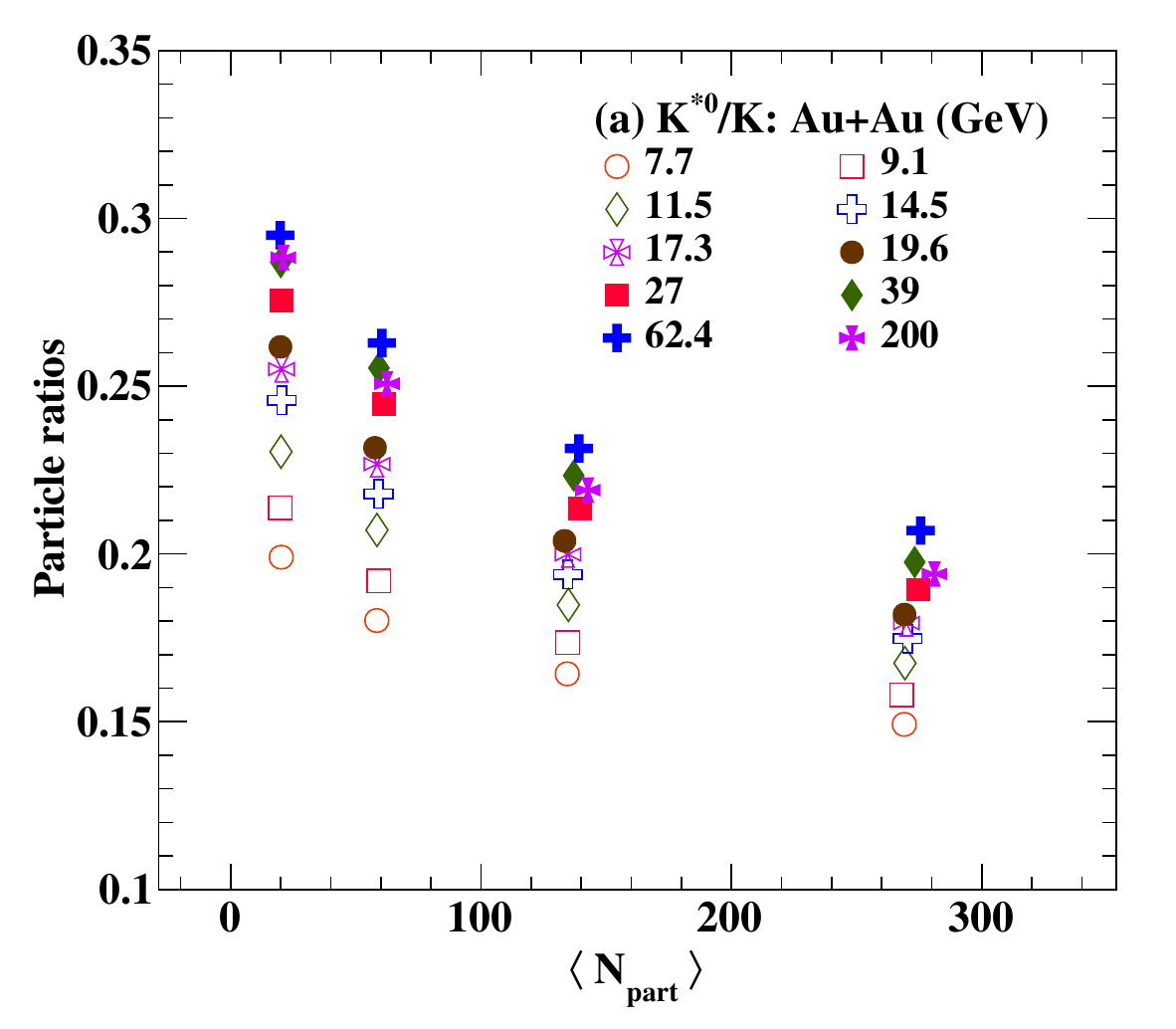}
        \end{minipage}
        \begin{minipage}{0.9\linewidth}
            \centering
            \includegraphics[scale=0.4]{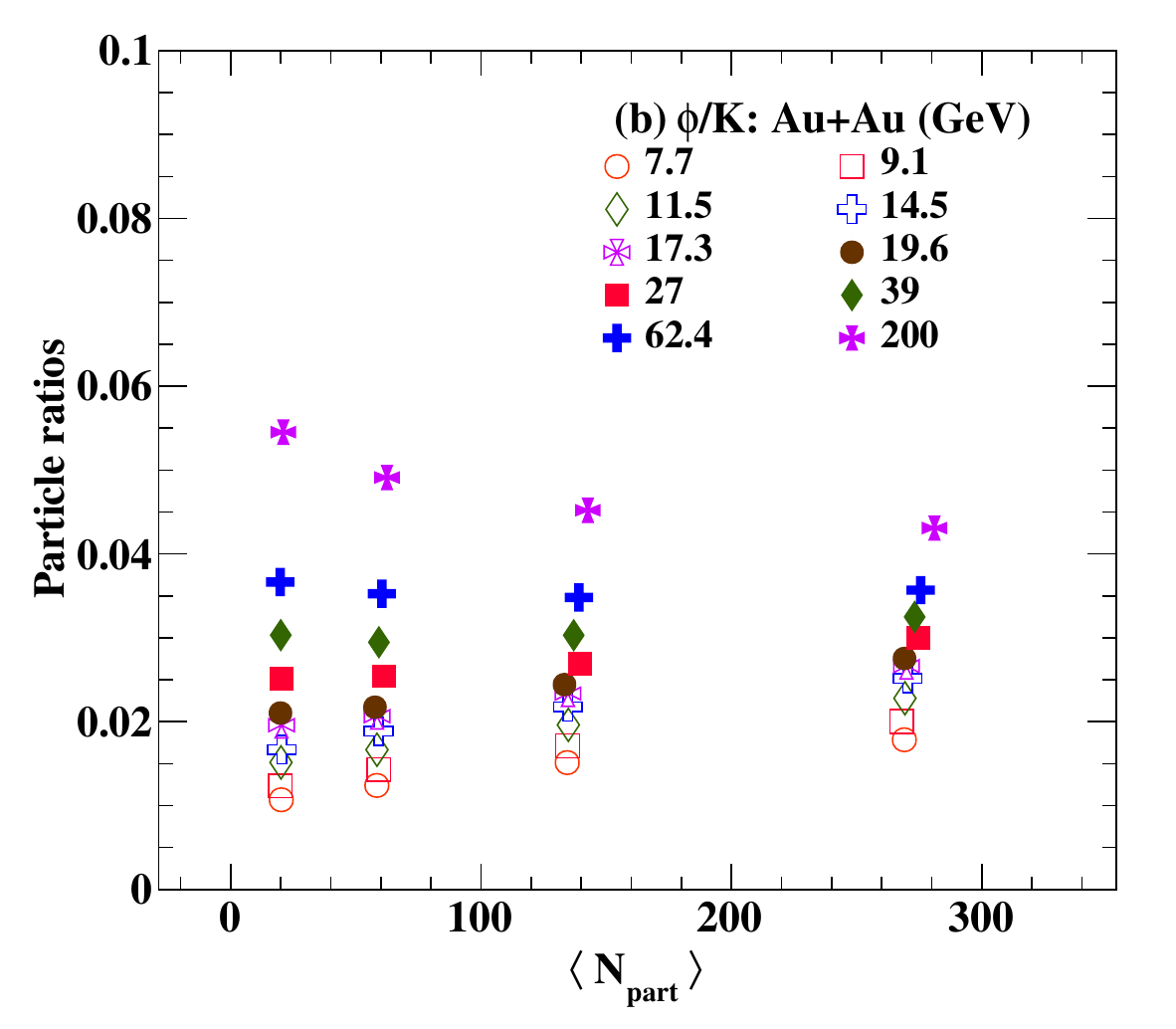}
        \end{minipage}
        \begin{minipage}{0.9\linewidth}
            \centering
            \includegraphics[scale=0.4]{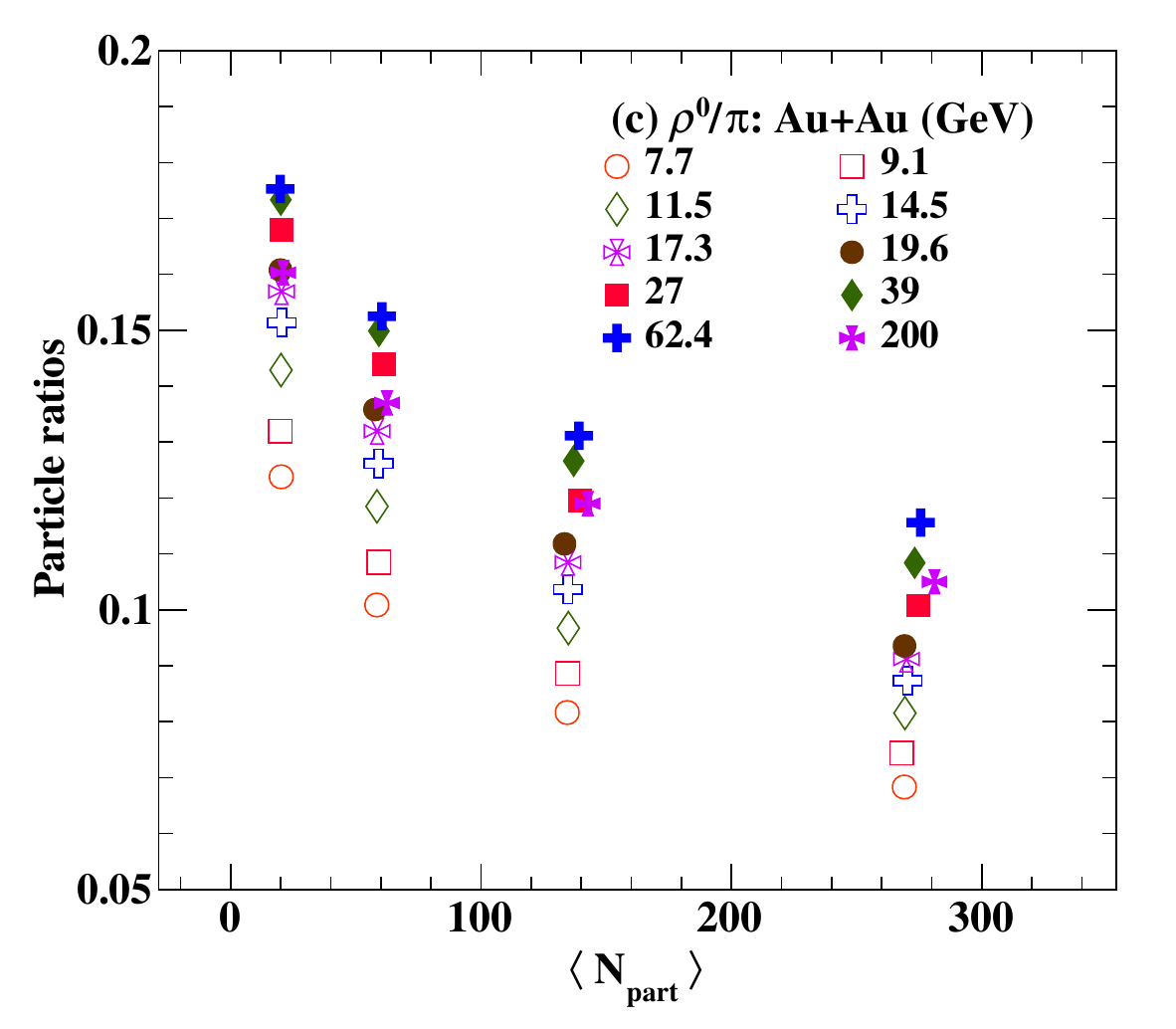}
        \end{minipage}  
		\caption{\label{fig:ratio_Npart}The yield ratios of (a) $K^{*0}/K$, (b) $\phi/K$ and (c) $\rho^{0}/\pi$ at mid-rapidity as a function of $\expval{N_\mathrm{part}}$ in Au+Au collisions from UrQMD at $\sqrt{s_\mathrm{NN}} = 7.7$, $9.1$, $11.5$, $14.5$, $17.3$, $19.6$, $27$, $39$, $62.4$ and $200$ GeV.}		
	\end{figure}

    The yield ratios of resonances to non-resonances serve as valuable observables for probing the hadronic rescatterings. Figure \ref{fig:ratio_sNN} presents the ratios of $K^{*0}/K \qty[=\qty(K^{*0}+\overline{K}^{*0})/\qty(K^{+}+K^{-})]$, $\phi/K\qty[=2\phi/\qty(K^{+}+K^{-})]$ and $\rho^0/\pi\qty[=2\rho^0/\qty(\pi^{+}+\pi^{-})]$ as a function of $\sqrt{s_\mathrm{NN}}$ in semi-central Au+Au collisions. The UrQMD results (solid symbols) are compared with the corresponding experimental data (open symbols)~\cite{kstarbes1, kstar62.4_200gev, phi62.4_200gev, phi200gev, rho_ppAuAu_star}. The $K^{*0}/K$ ratio from UrQMD is in agreement with the experimental data within uncertainties. The $\phi/K$ ratio from UrQMD is lowered than data by a factor of about four. This is also seen in several previous studies~\cite{kstarbes1_urqmd,lowphi1,lowphi2,lowphi3,lowphi4,phi62.4_200gev}. This could be due to inappropriate parameters for $\phi$ production in the default UrQMD setting. However, the energy dependence show similar trend in UrQMD and in data. Since this study focuses on the hadronic rescattering effect of $\phi$ meson, being dominated by the scatterings of the kaon from its decay and pions in the medium, the low $\phi/K$ ratio has negligible effect on this study. The $\rho^{0}/\pi$ ratio is compared with measurements from $40-80\%$ Au+Au collisions at $\sqrt{s_\mathrm{NN}}=$ 200 GeV, with the UrQMD results remaining within the experimental uncertainties. Figure \ref{fig:ratio_Npart} shows the ratios of $K^{*0}/K$, $\phi/K$ and $\rho^0/\pi$ as a function of $\expval{N_\mathrm{part}}$ at collision energy between $\sqrt{s_\mathrm{NN}}=$ 7.7 GeV and 200 GeV. The ratios decrease with increasing system size for $K^{*0}$ and $\rho^0$ at all energies, while the $\phi/K$ ratio exhibits no clear centrality dependence. All of these trends are consistent with experimental observations~\cite{kstarbes1, kstar62.4_200gev, kstar_phi2.76tev, kstar_phi2.76tev_rescattering, phi62.4_200gev, phi200gev, rho_2.76tev_PbPb}, indicating that the rescattering effects on vector mesons are reasonably well implemented in the UrQMD model at RHIC energies. Meanwhile, compared to $K^{*0}$ and $\rho^0$, the centrality dependence of $\phi$ is relatively weak, suggesting that hadronic interactions have a minimal effect on its production. In conclusion, UrQMD could effectively simulate resonance production and hadronic rescattering in heavy-ion collisions.
    
	\begin{figure*}[htbp]
		\centering
        \begin{minipage}{0.45\linewidth}
            \centering
            \includegraphics[scale=0.4]{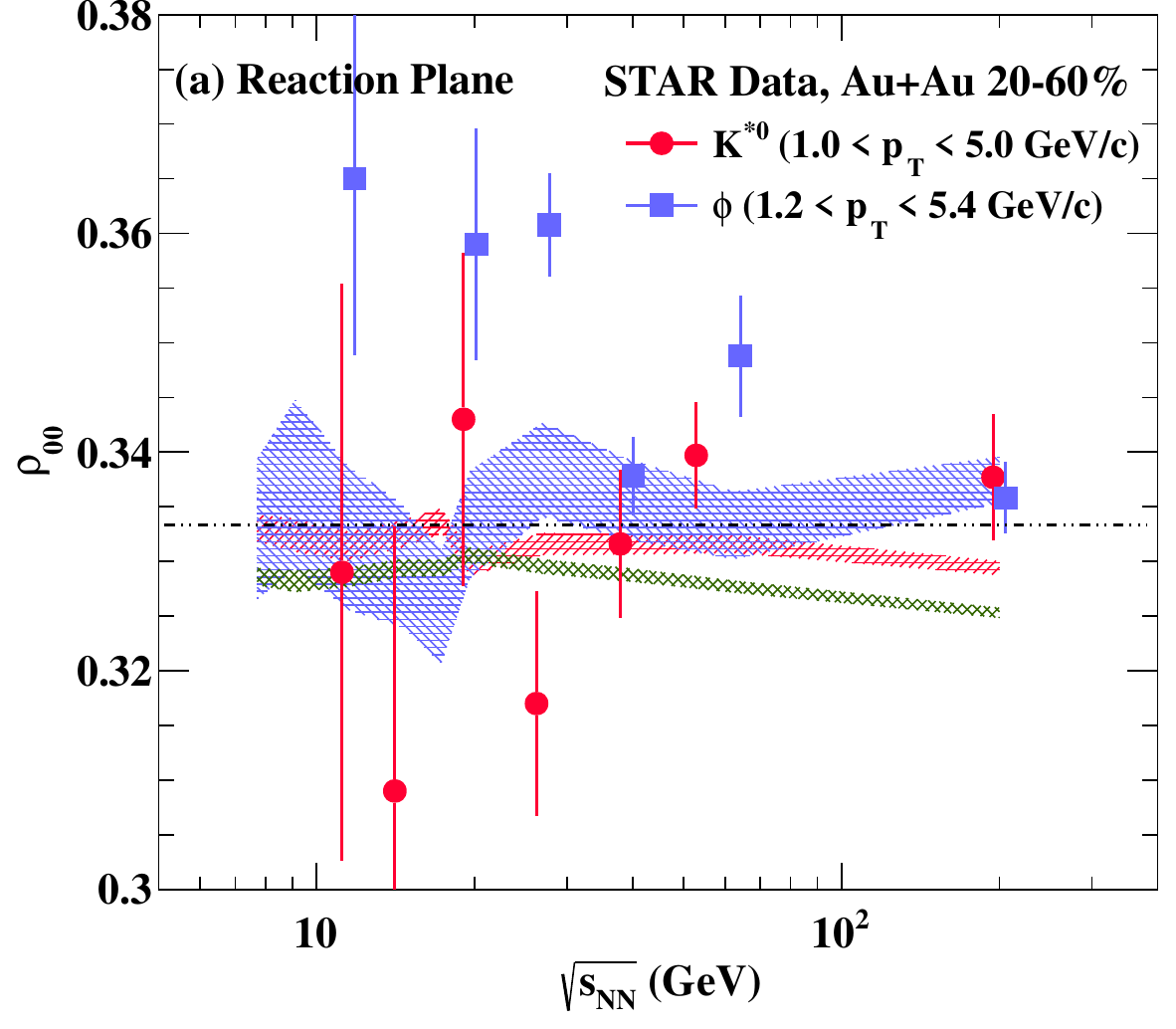}
        \end{minipage}
        \quad
        \begin{minipage}{0.45\linewidth}
            \centering
            \includegraphics[scale=0.4]{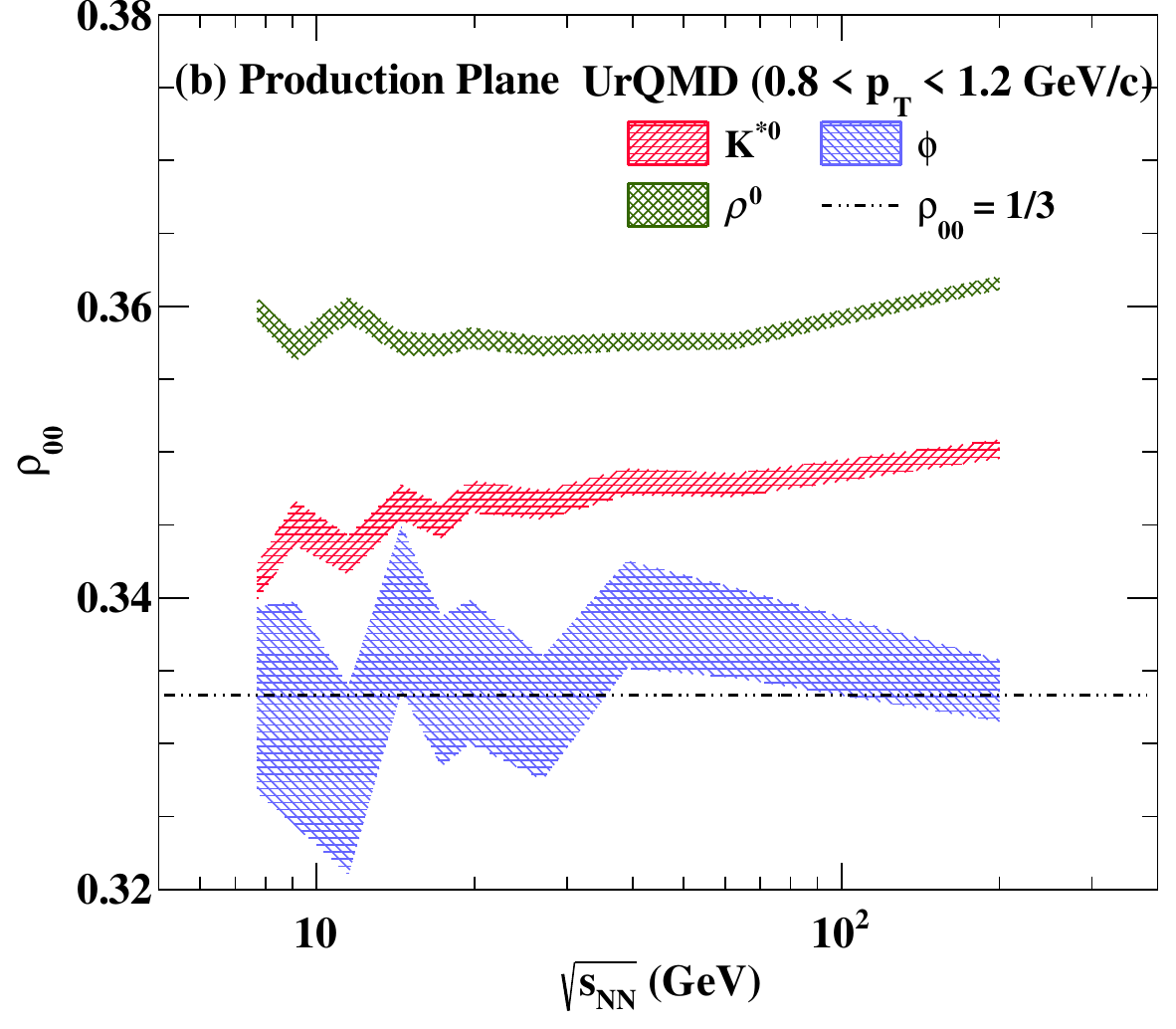}
        \end{minipage}
		\caption{\label{fig:rho00_RPPP}$\rho_{00}$ with respect to (a) reaction plane and (b) production plane for reconstructable $K^{*0}$, $\phi$ and $\rho^{0}$ for $0.8 < p_\mathrm{T} < 1.2 $ GeV/$c$ as a function of $\sqrt{s_\mathrm{NN}}$ in $20-60\%$ centrality Au+Au collisions from UrQMD. For $K^{*0}$ and $\phi$, the results are compared with previous STAR measurements~\cite{kstarphi_sa_star}, whose vertical bars represent summed errors.}
	\end{figure*}
 
	\subsection{Spin alignment with respect to reaction plane}
    


    The $\rho_{00}$ values for the initially generated vector mesons are verified to be consistent with $1/3$~\cite{kstar_spin_rescatterin}, the following discussions focus solely on reconstructable mesons. The left panel of \fig{fig:rho00_RPPP} shows $\rho_{00}$ for reconstructable $K^{*0}$, $\phi$ and $\rho^{0}$ mesons as a function of $\sqrt{s_\mathrm{NN}}$ for $ 0.8 < p_\mathrm{T} < 1.2$ GeV/$c$ in 20-60$\%$ Au+Au collisions with respect to reaction plane. The UrQMD results (hatched bands) are compared with the corresponding experimental data from STAR (filled circles and squares)~\cite{kstarphi_sa_star}. Since the reaction plane cannot be measured directly in the experiment, the STAR Collaboration reports $\rho_{00}$ values with respect to event plane, which serves as an estimation of the reaction plane. At all RHIC energies, the UrQMD results indicate that the $\rho_{00}$ values for $\rho^0$ are significantly below 1/3, whereas those for $\phi$ remain consistent with 1/3 within uncertainties. For $K^{*0}$, $\rho_{00}$ values show a clear deviation from 1/3 at higher energies but tend to approach 1/3 at lower energies. The negative values of $\rho_{00} - 1/3$ for $K^{*0}$ and $\rho^0$ arise from hadronic rescatterings in the anisotropic medium~\cite{kstar_spin_rescatterin}. The deviation of $K^{*0}$ from 1/3 at high energies is smaller than that of $\rho^{0}$, but larger than that of $\phi$. This pattern may be interpreted by considering that resonances with shorter lifetimes are more likely to decay within the dense medium, thereby increasing the probability that their decay daughters undergo rescattering. In addition, $\rho^0$ decays into two pions, making it more susceptible to rescattering effects than $K^{*0}$, which decays into a kaon and a pion.
    
    It also shows that $\rho_{00}$ values for $K^{*0}$ and $\rho^{0}$ tend to deviate more significantly from 1/3 as $\sqrt{s_\mathrm{NN}}$ increases. For reconstructable $K^{*0}$, $\rho_{00} - 1/3$ is $-0.0003\pm0.0012$ at $\sqrt{s_\mathrm{NN}}=$ 7.7 GeV, which is consistent with zero within uncertainties. However, at $\sqrt{s_\mathrm{NN}}=$ 200 GeV, $\rho_{00} - 1/3$ becomes $-0.0039 \pm 0.0007$, significantly below zero ($>5\sigma$). A similar trend is observed for reconstructable $\rho^0$, with $\rho_{00} - 1/3$ also decreasing with increasing $\sqrt{s_\mathrm{NN}}$ and reaching $-0.0080 \pm 0.0005$ ($\sim16\sigma$) at $\sqrt{s_\mathrm{NN}} = 200$ GeV. The growing deviation of $\rho_{00}$ from 1/3 for both reconstructable $K^{*0}$ and $\rho^{0}$ attributes to the increasing rescattering probability resulting from the higher medium density at higher energy. These large deviations from the UrQMD model suggest that the rescattering effect can significantly decrease the measured $\rho_{00}$ values for short-lived $K^{*0}$ and $\rho^{0}$, and such an effect on the measurement of spin alignment should be considered when comparing experimental measurements and theoretical predictions.
    
    

    \begin{figure*}[htbp!]
        \centering
        \begin{minipage}{0.45\linewidth}
            \centering
            \includegraphics[scale=0.4]{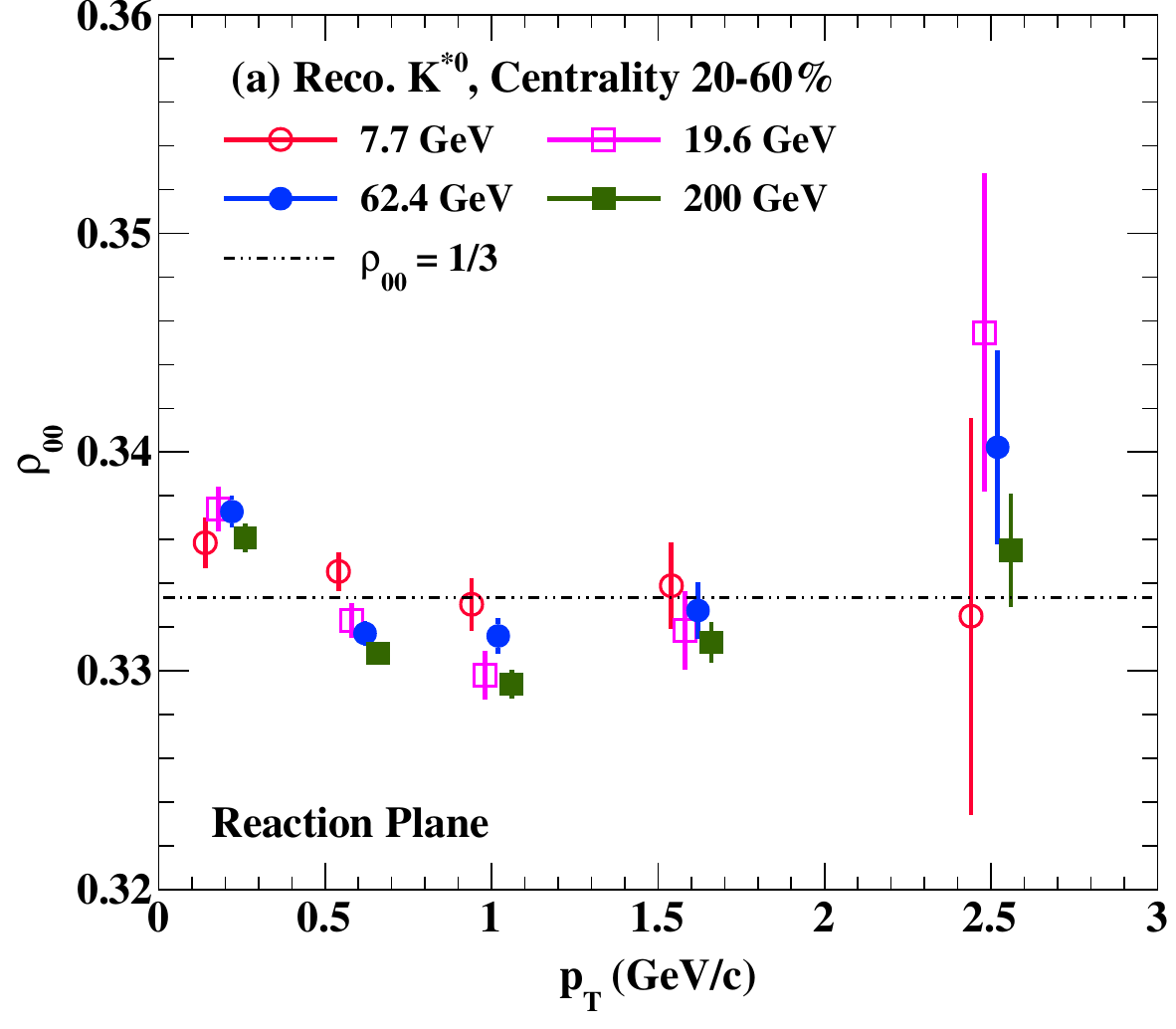}
        \end{minipage}
        \quad
        \begin{minipage}{0.45\linewidth}
            \centering
            \includegraphics[scale=0.4]{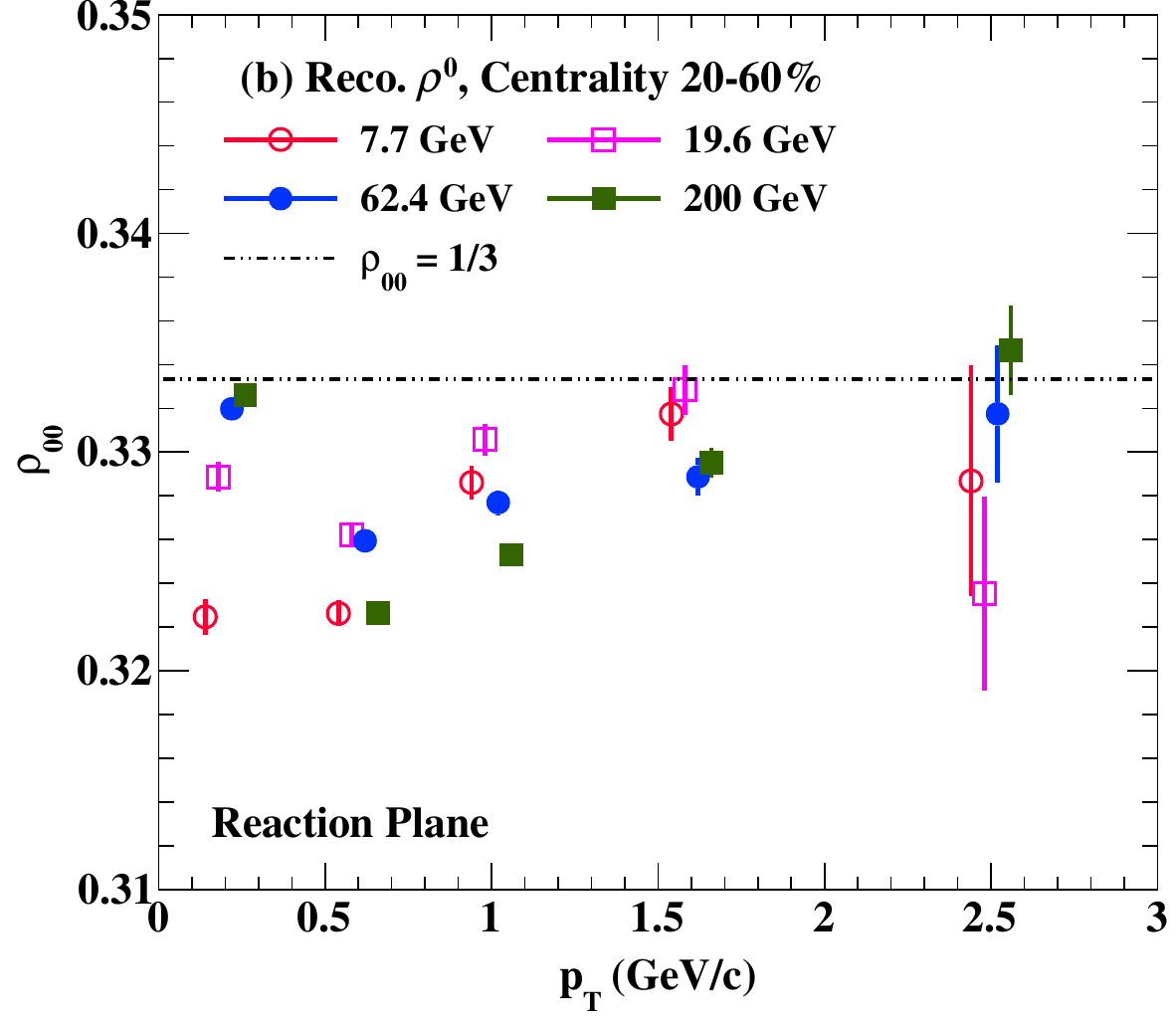}
        \end{minipage}
        \caption{\label{fig:rho00_RP_pT}$\rho_{00}$ with respect to reaction plane for reconstructable (a) $K^{*0}$ and (b) $\rho^{0}$ as a function of $p_\mathrm{T}$ in $20-60\%$ centrality in Au+Au collisions at $\sqrt{s_\mathrm{NN}} = 7.7$, $19.6$, $62.4$ and $200$ GeV.}       
    \end{figure*}
    
    \begin{figure*}[htbp!]
        \centering
        \begin{minipage}{0.45\linewidth}
            \centering
            \includegraphics[scale=0.4]{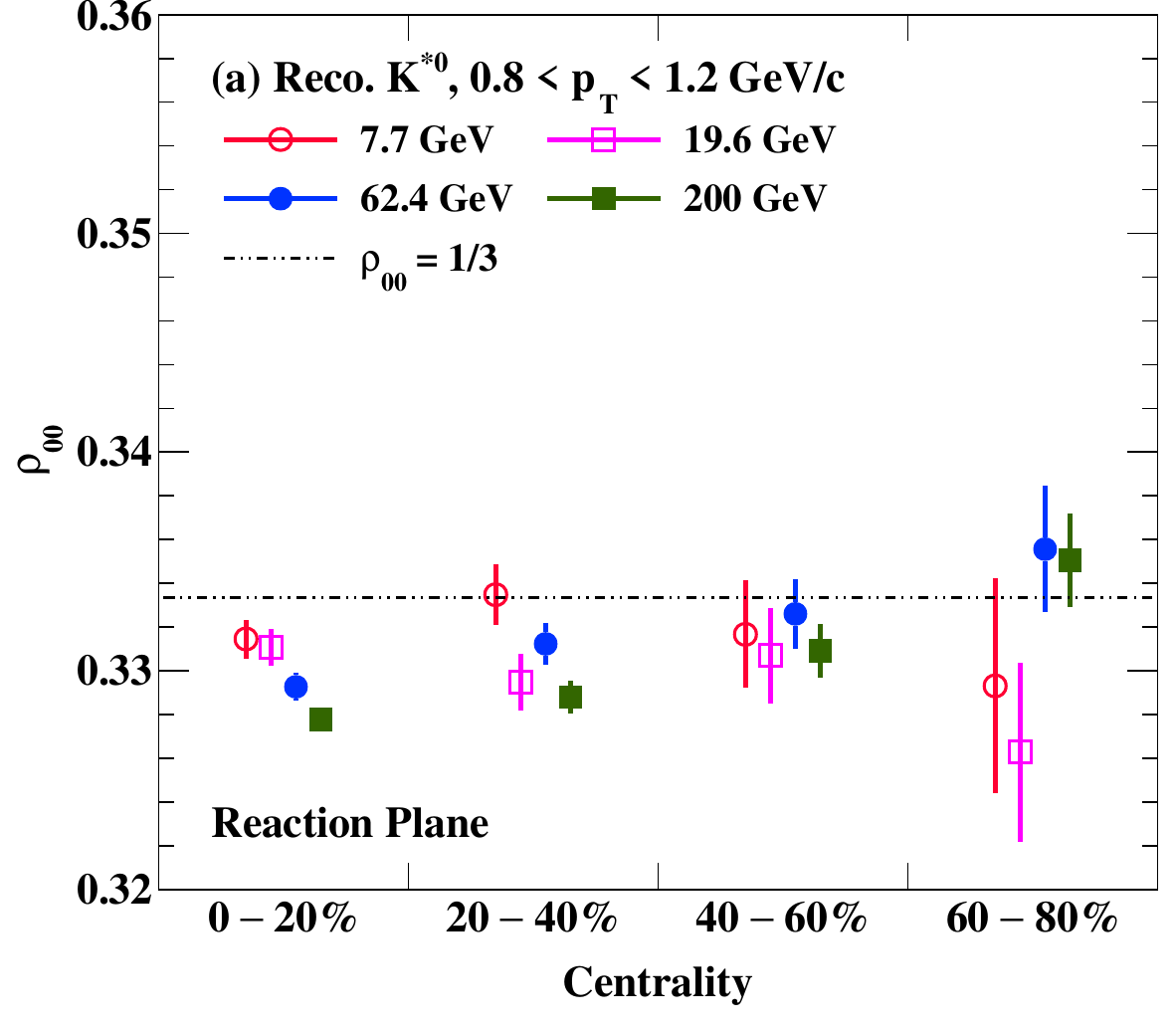}
        \end{minipage}
        \quad
        \begin{minipage}{0.45\linewidth}
            \centering
            \includegraphics[scale=0.4]{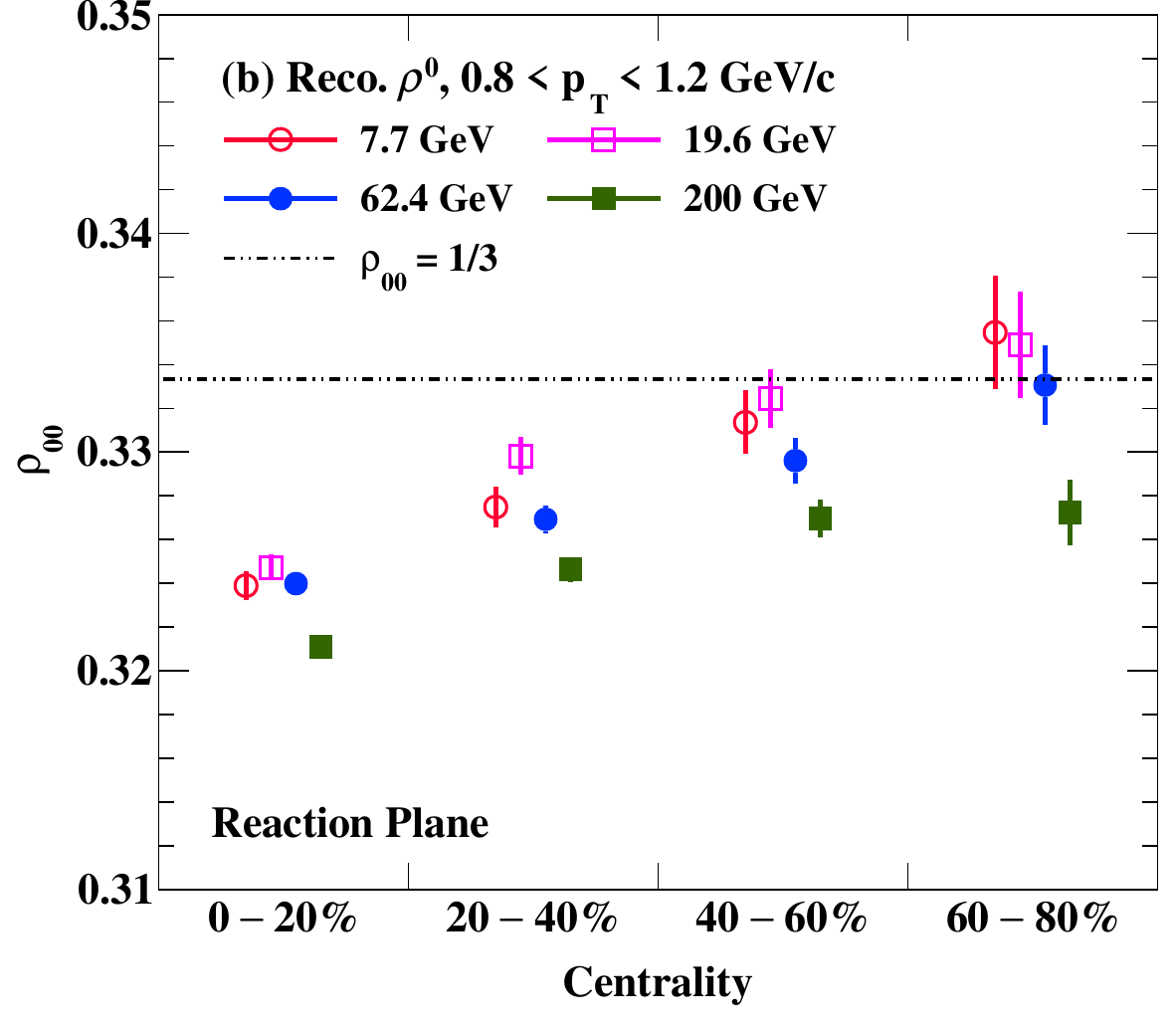}
        \end{minipage}
        \caption{\label{fig:rho00_RP_cen}$\rho_{00}$ with respect to reaction plane for reconstructable (a) $K^{*0}$ and (b) $\rho^{0}$ as a function of centrality for $0.8 < p_\mathrm{T} < 1.2 $ GeV/$c$ in Au+Au collisions at $\sqrt{s_\mathrm{NN}} = 7.7$, $19.6$, $62.4$ and $200$ GeV.}
    \end{figure*}
    
    
    \begin{figure*}[htbp]
        \centering
        \begin{minipage}{0.45\linewidth}
            \centering
            \includegraphics[scale=0.4]{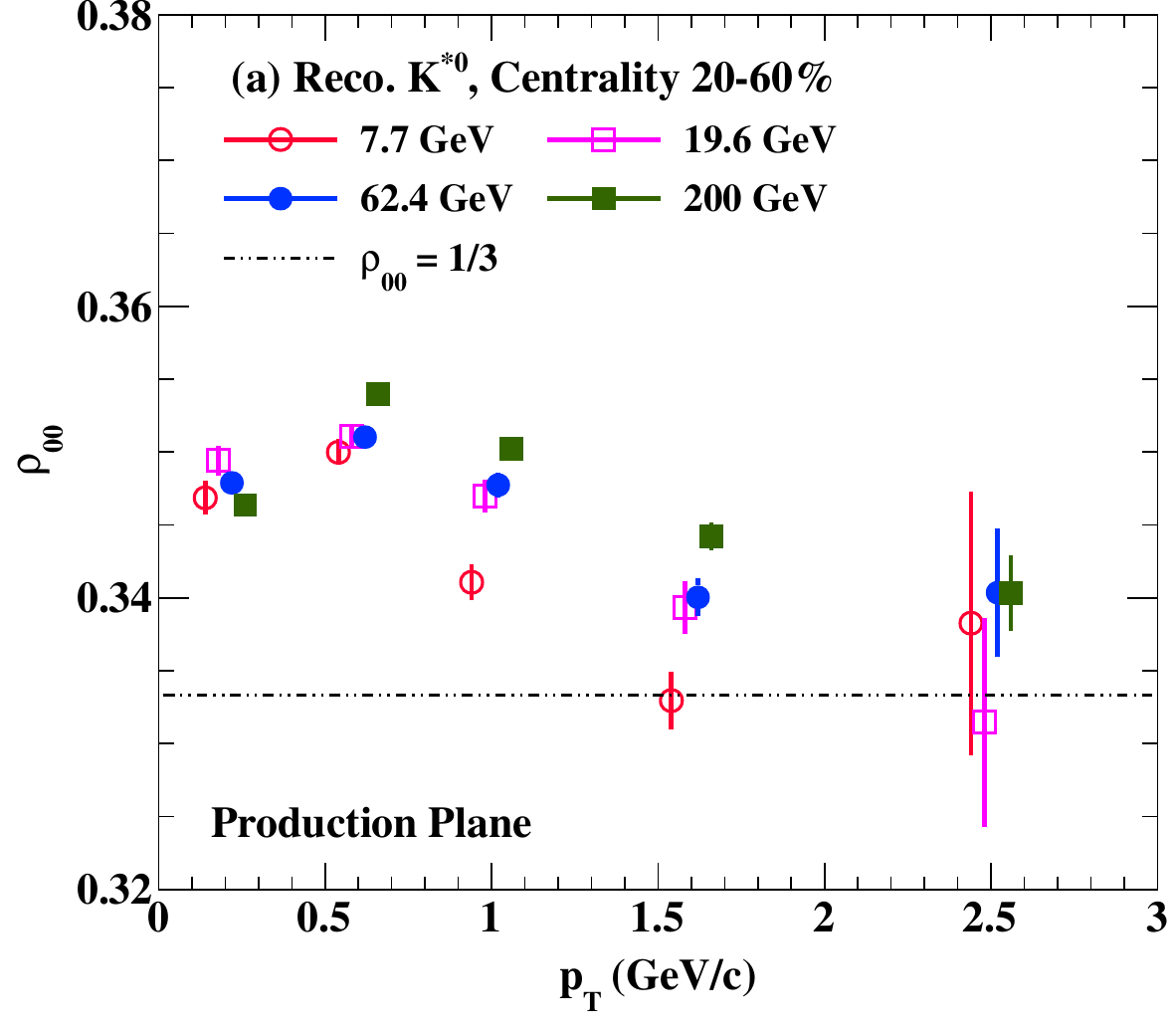}
        \end{minipage}
        \quad
        \begin{minipage}{0.45\linewidth}
            \centering
            \includegraphics[scale=0.4]{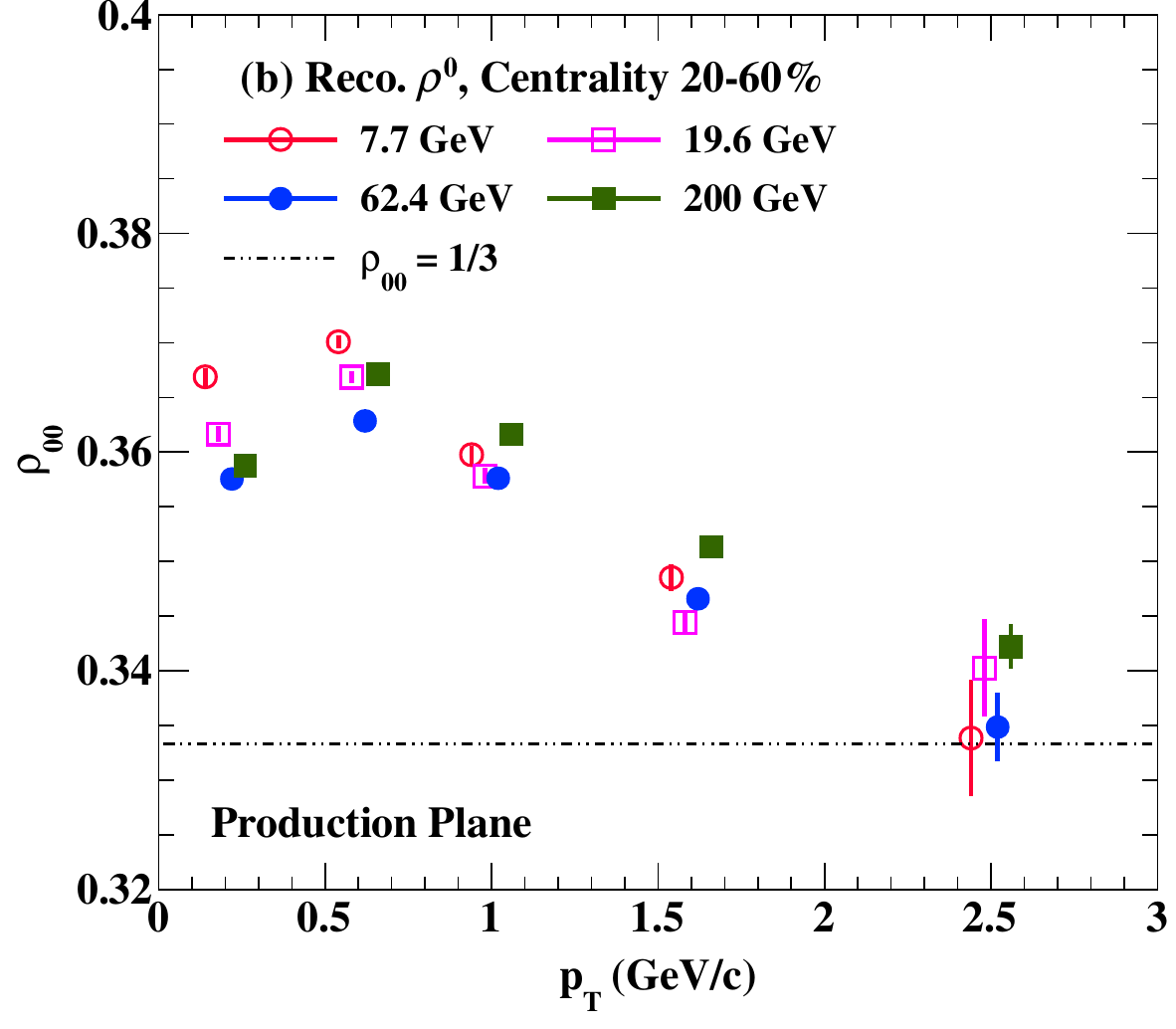}
        \end{minipage}
        \caption{\label{fig:rho00_PP_pT}$\rho_{00}$ with respect to production plane for reconstructable (a) $K^{*0}$ and (b) $\rho^{0}$ as a function of $p_\mathrm{T}$ in $20-60\%$ centrality in Au+Au collisions at $\sqrt{s_\mathrm{NN}} = 7.7$, $19.6$, $62.4$ and $200$ GeV.}
    \end{figure*}

    \begin{figure*}[htbp]
        \centering
        \begin{minipage}{0.45\linewidth}
            \centering
            \includegraphics[scale=0.4]{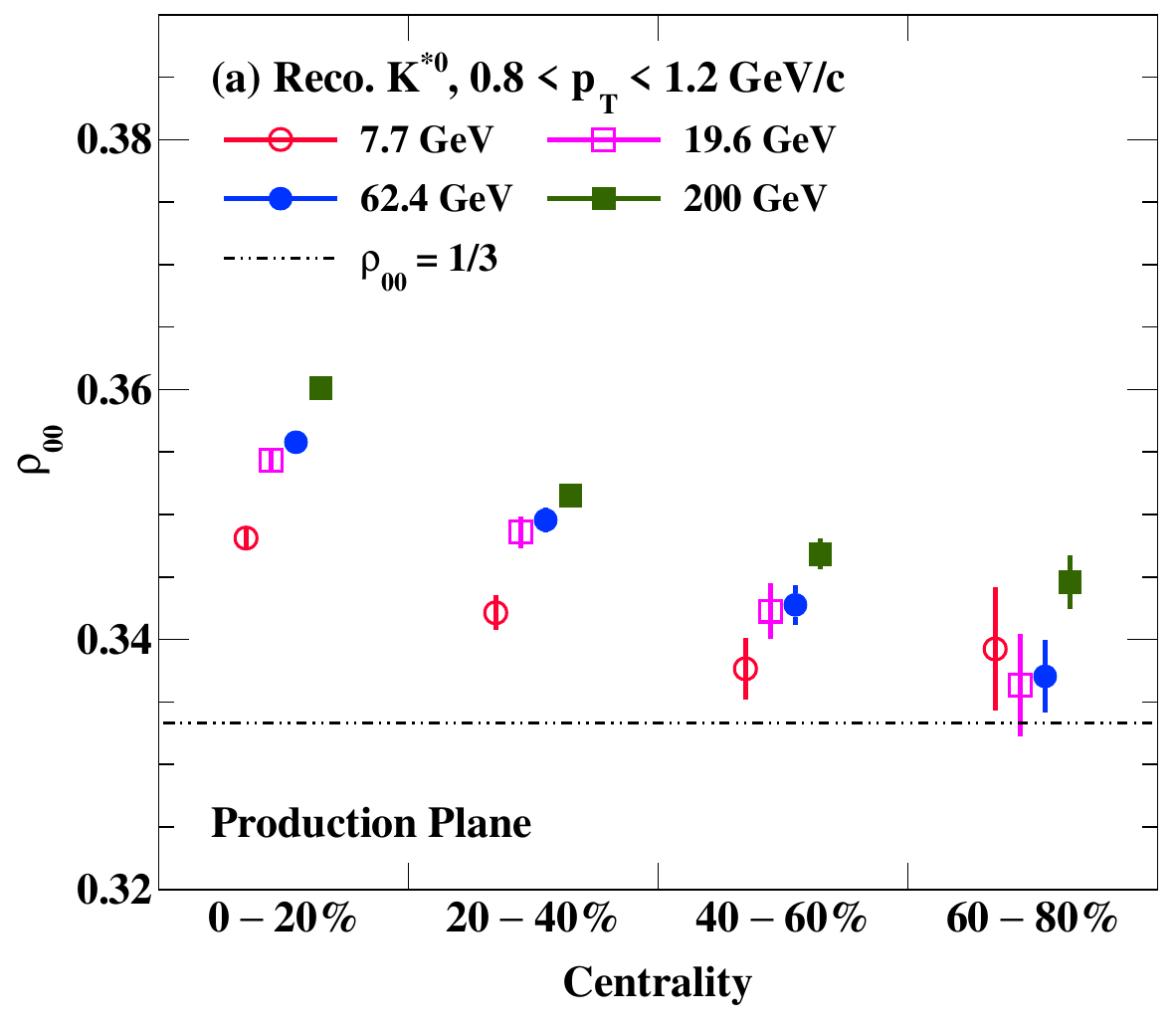}
        \end{minipage}
        \quad
        \begin{minipage}{0.45\linewidth}
            \centering
            \includegraphics[scale=0.4]{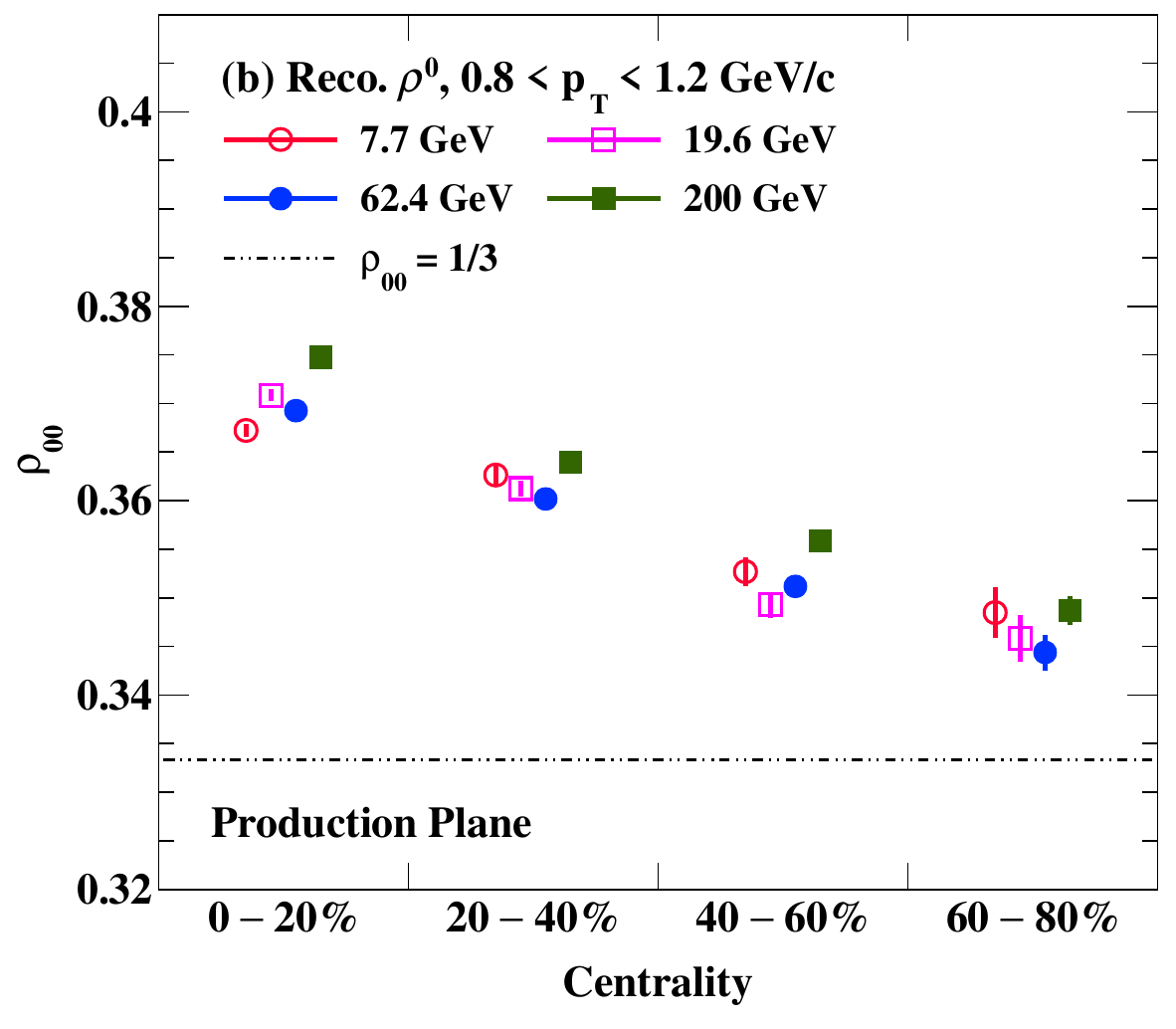}
        \end{minipage}
        \caption{\label{fig:rho00_PP_cen}$\rho_{00}$ with respect to production plane for reconstructable (a) $K^{*0}$ and (b) $\rho^{0}$ as a function of centrality for $0.8 < p_\mathrm{T} < 1.2 $ GeV/$c$ in Au+Au collisions at $\sqrt{s_\mathrm{NN}} = 7.7$, $19.6$, $62.4$ and $200$ GeV.}
    \end{figure*}

    As rescattering has no significant effect on the spin alignment of the $\phi$ meson, the following discussions focus exclusively on $\rho_{00}$ values of reconstructable $K^{*0}$ and $\rho^{0}$. Figure \ref{fig:rho00_RP_pT} presents the $p_\mathrm{T}$ dependence of $\rho_{00}$ with respect to reaction plane in 20-60\% Au+Au collisions at selected energies. Deviations of $\rho_{00}$ from 1/3 are observed for reconstructable $K^{*0}$ and $\rho^0$, with the largest deviations appearing in the intermediate $p_\mathrm{T}$ region. In the intermediate $p_\mathrm{T}$ region ($0.8<p_\mathrm{T}<1.2~\textrm{GeV}/c$), $\rho_{00}$ values for both $K^{*0}$ at high energies and $\rho^{0}$ are lower than 1/3. This is within expectations from the anisotropy of the medium. The deviation of $\rho_{00}$ values from 1/3 is considerably more pronounced for $\rho^0$ across all energies, suggesting a stronger rescattering effect on its spin alignment measurement. At low $p_\mathrm{T}$ ($p_\mathrm{T}<$ 0.4 GeV/$c$), the $\rho_{00}$ values for reconstructable $K^{*0}$ exceed 1/3. Such a trend is also seen for $\rho^0$ at high energies. One possible explanation for this increase at low $p_T$ could be related to the anisotropy $(v_{2})$ of the final hadrons in momentum space, which indicates more particle emission in reaction plane and consequently may increases the probability of rescattering in that direction, resulting in $\rho_{00} > 1/3$. In the high $p_\mathrm{T}$ region ($p_\mathrm{T}>$ 2.0 GeV/$c$), $\rho_{00}$ values for both $K^{*0}$ and $\rho^{0}$ are consistent with 1/3 within uncertainties due to reduced rescattering probability of their decay daughters.

  Figure \ref{fig:rho00_RP_cen} presents the centrality dependence of $\rho_{00}$ with respect to reaction plane for $K^{*0}$ and $\rho^{0}$ at $0.8<p_\mathrm{T}<1.2~\textrm{GeV}/c$ in Au+Au collisions at selected energies. $\rho_{00}$ values exhibit a clear centrality dependence for reconstructable $K^{*0}$ at higher energies and for $\rho^{0}$ at all energies. For both $K^{*0}$ and $\rho^{0}$, $\rho_{00}$ values exhibit their largest deviation from 1/3 in the most central collisions. The $\rho_{00} - 1/3$ values reach as large as $-0.0056 \pm 0.0005$ and $-0.0122 \pm 0.0004$ for $K^{*0}$  and $\rho^{0}$, respectively, at $\sqrt{s_\mathrm{NN}}=$ 200 GeV in 0-20\% central collisions, and gradually approach 1/3 in peripheral collisions. 


    \subsection{Spin alignment with respect to production plane}

    Following the investigation of spin alignment for these vector mesons with respect to reaction plane, $\rho_{00}$ values with respect to production plane are also studied. Similarly, the $\rho_{00}$ values as a function of $\sqrt{s_\mathrm{NN}}$ are investigated in the intermediate $p_\mathrm{T}$ region in 20-60\% Au+Au collisions, as shown in the right panel of \fig{fig:rho00_RPPP}. For reconstructable $\phi$, $\rho_{00}$ values remain consistent with 1/3. As the UrQMD model predicts a deviation of $\rho_{00}$ from 1/3 with respect to reaction plane for reconstructable $K^{*0}$ and $\rho^{0}$, a corresponding deviation is also expected with respect to production plane. This expectation is also confirmed by the UrQMD results. For both reconstructable $K^{*0}$ and $\rho^{0}$, $\rho_{00}$ values deviate from 1/3. Moreover, now this deviation from 1/3 is more significant compared to the values with respect to reaction plane. In particular, $\rho_{00}$ values for reconstructable $K^{*0}$ and $\rho^{0}$ are larger than 1/3 with respect to production plane. This is because daughters moving opposite to the direction of vector mesons in the rest frame are likely to be scattered due to the low momentum after Lorentz boost, resulting in $\rho_{00} > 1/3$. Such a significant difference between different reference planes has also been observed in previous studies~\cite{kstar_spin_rescatterin}. It can also be observed that the deviations of $\rho_{00}$ values for reconstructable $K^{*0}$ exhibit an increasing trend with rising $\sqrt{s_\mathrm{NN}}$. $\rho_{00} - 1/3$ values increase from $0.0077 \pm 0.0012$ at $\sqrt{s_\mathrm{NN}}=$ 7.7 GeV to $0.0169 \pm 0.0007$ at $\sqrt{s_\mathrm{NN}}=$ 200 GeV. For reconstructable $\rho^{0}$, $\rho_{00}$ shows weak dependence on energy at low energy ($\sqrt{s_\mathrm{NN}}< 62$ GeV), and increasing tread at higher energy.The largest $\rho_{00} - 1/3$ reaches $0.0283 \pm 0.0005$ at 200 GeV.
    
    Figure \ref{fig:rho00_PP_pT} presents the $p_\mathrm{T}$ dependence for 20-60\% centrality and \fig{fig:rho00_PP_cen} shows the centrality dependence in the intermediate $p_\mathrm{T}$ region ($ 0.8 < p_\mathrm{T} < 1.2$ GeV/$c$). The deviation of $\rho_{00}$ from 1/3 with respect to production plane is significant at low $p_\mathrm{T}$ and in central collisions, while $\rho_{00}$ values become consistent with 1/3 at high $p_\mathrm{T}$ and in peripheral collisions. $\rho_{00}$ values decrease quickly towards peripheral collisions, because the rescattering effect is more dominant in central collisions than in peripheral heavy-ion collisions. Notably, the maximum deviation of $\rho_{00}$ from 1/3 occurs in the intermediate $p_\mathrm{T}$ region and central collisions. The $\rho_{00} - 1/3$ values reach up to $0.0268\pm0.0005$ and $0.0414\pm0.0004$ for $K^{*0}$ and $\rho^{0}$, respectively.
	
	\section{\label{section:4}Summary}

    In summary, we have studied the rescattering effect on the measurement of $\rho_{00}$ for $K^{*0}$, $\phi$ and $\rho^0$ mesons in Au+Au collisions using the UrQMD model, with collision energies covering the BES and top RHIC energies. The $\rho_{00}$ values are extracted with respect to both reaction plane the production plane, and compared with the STAR data. For reconstructable $\phi$, the $\rho_{00}$ values exhibit no significant deviation from 1/3, implying that rescattering has a minimal impact on the $\rho_{00}$ measurements of long-lived vector mesons. In contrast, the $\rho_{00}$ values for reconstructable $K^{*0}$ and $\rho^{0}$ appear significant deviations from 1/3 with respect to both planes, suggesting a pronounced rescattering effect on the $\rho_{00}$ measurements of short-lived vector mesons. The deviations of $\rho_{00}$ from 1/3 show an increasing trend with increasing $\sqrt{s_\mathrm{NN}}$ for both reconstructable $K^{*0}$ and $\rho^{0}$. The dependencies of $\rho_{00}$ on $p_\mathrm{T}$ and centrality show strongest rescattering effect in intermediate $p_\mathrm{T}$ and in central collisions. The maximum deviation reaches $-0.0056$ ($0.0268$) for $K^{*0}$ and $-0.0122$ ($0.0414$) for $\rho^{0}$ with respect to the reaction (production) plane. Such large deviations should be taken into account in the comparison between experimental measurements and theoretical predictions.
    
    
\section*{Acknowledgment}
    This work is supported in part by National Natural Science Foundation of China (NSFC) under Contract Nos. 1231101148 and 12175223, and National Key Research and Development Program of China under Contract No. 2022YFA1604900.
    
    \bibliography{references.bib}

\end{document}